\documentclass[journal]{IEEEtran}
\usepackage[nolist,nohyperlinks]{acronym}
\usepackage{graphics}
\usepackage[pdftex]{graphicx}
\usepackage{subcaption}
\usepackage{xcolor}
\usepackage{hyperref}
\colorlet{corr1}{black}
\colorlet{corr2}{black}
\colorlet{corr3}{black}

\usepackage{cite}

\usepackage{soul,color}

\usepackage{multirow}
\usepackage{amsmath}
\usepackage{mwe}

\begin{document}
%
\title{ML-based handover prediction over a real O-RAN deployment using RAN Intelligent controller}
%
%
%

\author{Merim~Dzaferagic\IEEEauthorrefmark{1}, Bruno~Missi~Xavier\IEEEauthorrefmark{2}, Diarmuid~Collins\IEEEauthorrefmark{1}, Vince~D'Onofrio\IEEEauthorrefmark{3}, Magnos~Martinello\IEEEauthorrefmark{2}
        and~Marco~Ruffini\IEEEauthorrefmark{1}\\
	\IEEEauthorblockA{\IEEEauthorrefmark{1}CONNECT Research Centre, School of Computer Science and Statistics, Trinity College Dublin\\}
	\IEEEauthorblockA{\IEEEauthorrefmark{2}Federal University of Espirito Santo, Espirito Santo, Brazil\\}
	\IEEEauthorblockA{\IEEEauthorrefmark{3}Rivada Networks\\}
	Emails: dzaferam@tcd.ie, bruno.xavier@ifes.edu.br, dcollin5@tcd.ie, vdonofrio@rivada.com, magnos.martinello@ufes.br and marco.ruffini@tcd.ie}
\maketitle

\begin{abstract}
O-RAN introduces intelligent and flexible network control in all parts of the network. The use of controllers with open interfaces allow us to gather real time network measurements and make intelligent/informed decision. The work in this paper focuses on developing a use-case for open and reconfigurable networks to investigate the possibility to predict handover events and understand the value of such predictions for all stakeholders that rely on the communication network to conduct their business. We propose a Long-Short Term Memory Machine Learning approach that takes standard Radio Access Network measurements to predict handover events. The models were trained on real network data collected from a commercial O-RAN setup deployed in our OpenIreland testbed. Our results show that the proposed approach can be optimized for either recall or precision, depending on the defined application level objective. We also link the performance of the Machine Learning (ML) algorithm to the network operation cost. Our results show that ML-based matching between the required and available resources can reduce operational cost by more than $80\%$, compared to long term resource purchases. 
\end{abstract}


%
\IEEEpeerreviewmaketitle

\begin{acronym}
\acro{fp}[FP]{False Positive}
\acro{fn}[FN]{False Negative}
\acro{tp}[TP]{True Positive}
\acro{v2x}[V2X]{Vehicle-to-Everything}
\acro{ai}[AI]{Artificial Intelligence}
\acro{ml}[ML]{Machine Learning}
\acro{ric}[RIC]{RAN Intelligent Controller}
\acro{non-rt}[non-RT]{non-Real Time}
\acro{near-rt}[near-RT]{near-Real Time}
\acro{rt}[RT]{real-time}
\acro{rru}[RRU]{Radio Remote Unit}
\acro{bs}[BS]{Base Station}
\acro{vbs}[vBS]{Virtual Base Station}
\acro{mcs}[MCS]{Modulation and Coding Scheme}
\acro{cqi}[CQI]{Channel Quality Indicator}
\acro{nn}[NN]{Neural Network}
\acro{snr}[SNR]{Signal to Noise Ratio}
\acro{ran}[RAN]{Radio Access Network}
\acro{oran}[O-RAN]{Open Radio Access Network}
\acro{nr}[NR]{New Radio}
\acro{5g}[5G]{Fifth Generation}
\acro{3gpp}[3GPP]{3rd Generation Partnership Project}
\acro{cu}[CU]{Central Unit}
\acro{du}[DU]{Distributed Unit}
\acro{ru}[RU]{Radio Unit}
\acro{smo}[SMO]{Service Management and Orchestration}
\acro{iot}[IoT]{Internet of Things}
\acro{ott}[OTT]{Over-The-Top}
\acro{uav}[UAV]{Unmanned Aerial Vehicle}
\acro{qos}[QoS]{Quality of Service}
\acro{roadm}[ROADM]{Reconfigurable Optical Add-Drop Multiplexer}
\acro{gis}[GIS]{Geographic Information System}
\acro{lstm}[LSTM]{Long-Short Term Memory}
\acro{svm}[SVM]{Support Vector Machine}
\acro{ue}[UE]{User Equipment}
\acro{rsrp}[RSRP]{Reference Signal Received Power}
\acro{rsrq}[RSRQ]{Reference Signal Received Quality}
\acro{sinr}[SINR]{Signal-to-Interference \& Noise Ratio}
\acro{sla}[SLA]{Service Level Agreement}
\acro{re}[RE]{Resource Element}
\acro{rs}[RS]{Reference Signal}
\acro{rssi}[RSSI]{Received Signal Strength Indicator}
\acro{rnn}[RNN]{Recurrent Neural Network}
\acro{rb}[RB]{Resource Block}
\acro{kpi}[KPI]{Key Performance Indicator}
\acro{sm}[SM]{Service Model}
\acro{ie}[IE]{Information Element}
\end{acronym}
\section{Introduction}\label{sec:introduction}
\begin{figure}
    \centering
    \includegraphics[scale=0.7]{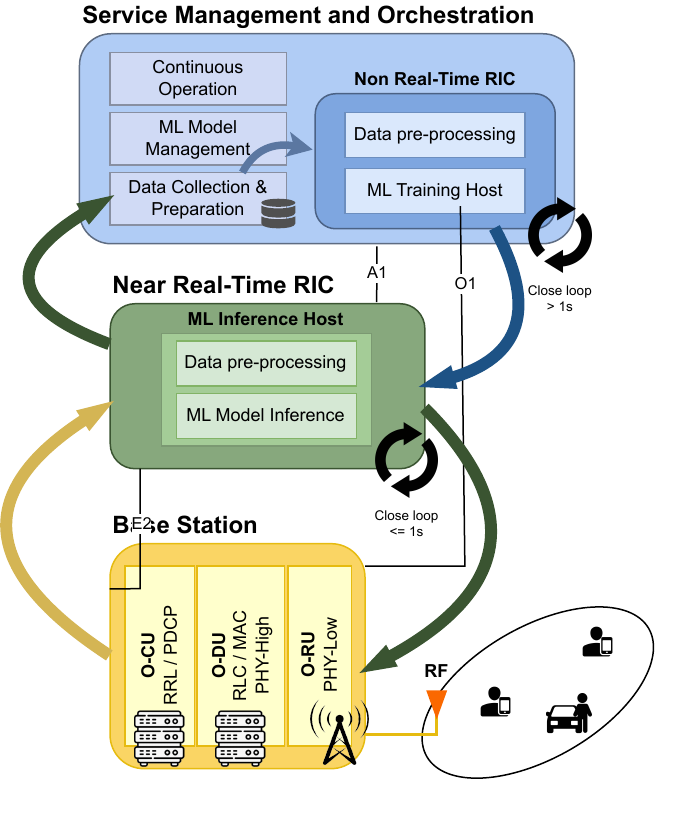}
    \caption{O-RAN architecture showing that the functional parts of the CU/DU can be moved closer to and further from the RU.}
    \label{fig:oran_architecture}
    \vspace{-1.5em}
\end{figure}

The \ac{ran} is moving towards a more open, disaggregated and service-oriented architecture. The main enablers of these changes are the shift from proprietary protocols and closed systems to commodity hardware and standardized interfaces, and the introduction of \ac{5g} \ac{nr} functional splits \cite{schmidt2021flexric}. Along these lines, as shown in Figure~\ref{fig:oran_architecture}, the \ac{3gpp} defined a new flexible architecture for the \ac{5g} \ac{ran} in which the \acp{bs} (i.e. gNodeBs) are split into three logical parts (i.e., \ac{cu}, \ac{du} and \ac{ru})\cite{3gppRelease14,3gpp2019nr}. Each of these parts is capable of hosting different functions of the \ac{5g} \ac{nr} stack. Further, \ac{3gpp} defines eight options for distributing the functionality of the \ac{nr} \ac{ran} stack across the fronthaul network (i.e. eight functional splits). The functional splits allow for different levels of distribution and centralization of network functions, which enables the exploitation of the trade-off between cooperation and coordination in the \ac{ran}\cite{rodriguez2020cloud,duan2016performance}. O-RAN focuses specifically on the 7.2x split \cite{3gpp2011evolved, polese2022understanding}.

The O-RAN paradigm is an effort to achieve high flexibility, intelligence, and openness in the next-generation of \acp{ran} \cite{oran,polese2022understanding}. Moreover, full interface specification allows different vendors' compatibility around an ecosystem of mobile networks. The intelligence is introduced by the \acp{ric} that provide \ac{rt}, near-\ac{rt} and non-\ac{rt} control loops \cite{alliance2019ran, alliance181017ran}. 
The \ac{rt} \ac{ric} is a logical function that enables control and optimization on a $10$ms timescale. The \ac{near-rt} \ac{ric}, enables control and optimization of O-RAN elements and resources via fine-grained data collection and actions over the \emph{E2} interface. The \ac{near-rt} \ac{ric} allows for a delay between $10$ms up to $1$s \cite{alliance2020ranG3}. Lastly, the \ac{non-rt} \ac{ric} \cite{alliance2021ranG2} allows for control loop delays above $1$s. With the control elements in place, we can leverage the inherent flexibility provided by the functional splits to extract valuable insights from the \ac{ran}. By capitalizing on the above-mentioned trade-off between cooperation and coordination, we can fine-tune the configuration to align with a wide range of objectives. For example, multiple \acp{ru} can be connected to one or more \acp{du}. The same holds for the connection between the \acp{du} and \acp{cu}. Different configurations allow us to tweak the trade-off between optimized central control/coordination and distributed computation and resilience. The configurations can be updated dynamically through the open interfaces. Figure~\ref{fig:oran_architecture} provides an overview of the existing control loops and \ac{ml} workflows that can be mapped onto them. The \ac{non-rt} \ac{ric} supports model training and updates, and policy-based guidance of applications in the \ac{near-rt} \ac{ric} \cite{alliance2019ran2}. It also supports data pre-processing, storage and \ac{ml} model performance monitoring over time. The \ac{near-rt} \ac{ric} is more constrained due to its time limitations, and therefore it only performs data collection, simple pre-processing and inference, or simple policy-based actions. 

The use of \ac{ml} within these control loops allows us to understand and enhance the network operation through intelligence embedded in the network architecture. The authors of \cite{alliance2019ran2} address the needed components and interfaces to design, train and deploy such \ac{ml} models. Recent publications show that the use of \ac{ric} applications provides a data-driven approach to network optimization \cite{dryjanski2021toward, johnson2021open}, which enables efficient resources allocation \cite{bashir2019optimal, zhang2022federated, zhang2022team}, improves the spectrum efficiency through intelligent resource orchestration at the edge \cite{9903911, ali2021multi}, and optimizes the network configuration for power consumption \cite{dzaferagic2022Globe}. The authors of \cite{bhattarai2016overview, ansari2020spectrum} highlight the scarcity of spectrum and the resulting high cost. They also highlight that dynamic spectrum sharing will significantly reduce the operational cost especially for small service providers, and improve the spectrum utilization. \textit{We focus on minimizing that cost by applying \ac{ml}-based decision making to resource acquisition. }Unlike the theoretical approach to discussing the potential of intelligent control aspects in O-RAN, we have built a testbed based on commercial grade hardware and software to test and study the effects that intelligent control has on the network operation. 

Considering the high mobility of \acp{ue} in wireless networks, their behavioral patterns, interference, and noise conditions, predicting the resource requirements in different coverage areas of the network remains a challenging task. Different approaches have been used to predict user movement and prepare the mobile network for classic handover triggers. 
In \cite{roy2004exploiting, demissie2013exploring}, the authors take a statistical approach to understand the \ac{qos} and continuity of communication services. The authors of \cite{roy2004exploiting} propose an Information-theoretic mobility management algorithm related to entropy and information content of the user’s movement to reduce the signaling overhead by around 80\%. The work relies on data collected from a simulation environment. The authors of \cite{demissie2013exploring} take advantage of \ac{gis} data to explore the handovers and analyze the mobility in urban centers. A Markov chain is proposed in \cite{yan2021mobility} for mobility prediction based on mobile user trajectory. The authors apply the weighting coefficients to the Markov prediction to improve the model's accuracy. The authors of \cite{9171421} rely on a distributed optimization algorithm that estimates user mobility patterns to configure the association between \acp{bs} and continues handover regions to minimize the signaling overhead required during the handover procedure. Unlike these approaches, \textit{our goal is to predict handovers on a higher granularity level (i.e. on a multi-second time scale), to allow dynamic resource acquisition for the purpose of network operation cost optimization. }

Furthermore, the authors of \cite{sun2020efficient} propose a multi-agent LEarning based Smart handover Scheme (LESS) to minimize the handover cost while ensuring the agreed user's \ac{qos} requirements. The authors of \cite{kaur2022efficient}, proposes a hybrid handover technique. They use \ac{lstm} and \ac{svm} algorithms to predict the parameters of a mobile device (e.g. location coordinates, speed, reference signal received power, and reference signal received quality). Instead of redesigning handover procedures, we focus on the prediction of those events that can be used by all network stakeholders to optimize various aspects of their network configuration (e.g. latency, resilience, resource utilization optimization). \textit{Our approach is agnostic to the handover techniques and \ac{qos} models and rather focuses on the handover prediction for the purpose of business level decisions related to resource acquisition. }

In \cite{cai2021mobility}, the authors use user mobility predictions to reduce backhaul traffic in wireless networks. To do this, they predict the user movement in a simulated environment, relying on \ac{lstm} in three different scenarios: (i) a linear movement, when the user walks in a straight line on the street; (ii) circular movement, simulating a fixed path trajectory, and; (iii) random movement, applying the irregular trajectory in an open area. The results show significant performance improvement in all scenarios, mainly when the mobility follows a deterministic movement. Unlike the work in \cite{cai2021mobility}, we use realistic data collected from the network deployed in our testbed. Additionally, instead of predicting user mobility patterns directly, we focus on the handover event prediction which allows us to extract information about the user mobility through generic signal measurements (e.g. \ac{rsrp}, \ac{rsrq} and \ac{sinr}).

\textit{In summary, unlike the majority of the work proposed in the literature our work relies on real data sets collected from a real O-RAN deployment in our testbed.} This approach allows us to avoid unrealistic assumptions about the availability of information used for various predictions, and allows us to study the potential of predicting handover events based only on the information that is available in the \ac{ran}. \textit{Additionally, instead of focusing on improving handover procedures or optimizing the network for \ac{qos}, our goal is to predict handover events for the purpose of network operation cost/performance optimization. Despite advancements in \ac{ml}-based mobility predictions and handover procedure optimization, there remains a gap in understanding how \ac{ml} performance metrics influence application-level requirements specified by the communication network.} We address this gap and provide details about the potential impact that such predictions can have on the different network stakeholders. 



The main contributions of this work include: 
\begin{itemize}
    \item We provide a detailed description of a testbed setup that combines open-source and commercial grade equipment, with licensed spectrum and capabilities to integrate edge and cloud and mirroring a wide variety of real-world infrastructure/functional topologies;
    \item We introduce and assess a use-case that examines the feasibility of a network control and optimization provider in gathering performance measurements and forecasting handover events;
    \item We propose a cost function for application level optimization that links the cost of resource acquisition in a dynamic network environment with \ac{ml} performance metrics (e.g. recall, precision).
\end{itemize}

\section{Problem Description and Experimental Setup}\label{sec:problem_description_and_experimental_setup}
\begin{figure*}[t!]
    \centering
    \includegraphics[scale=0.25]{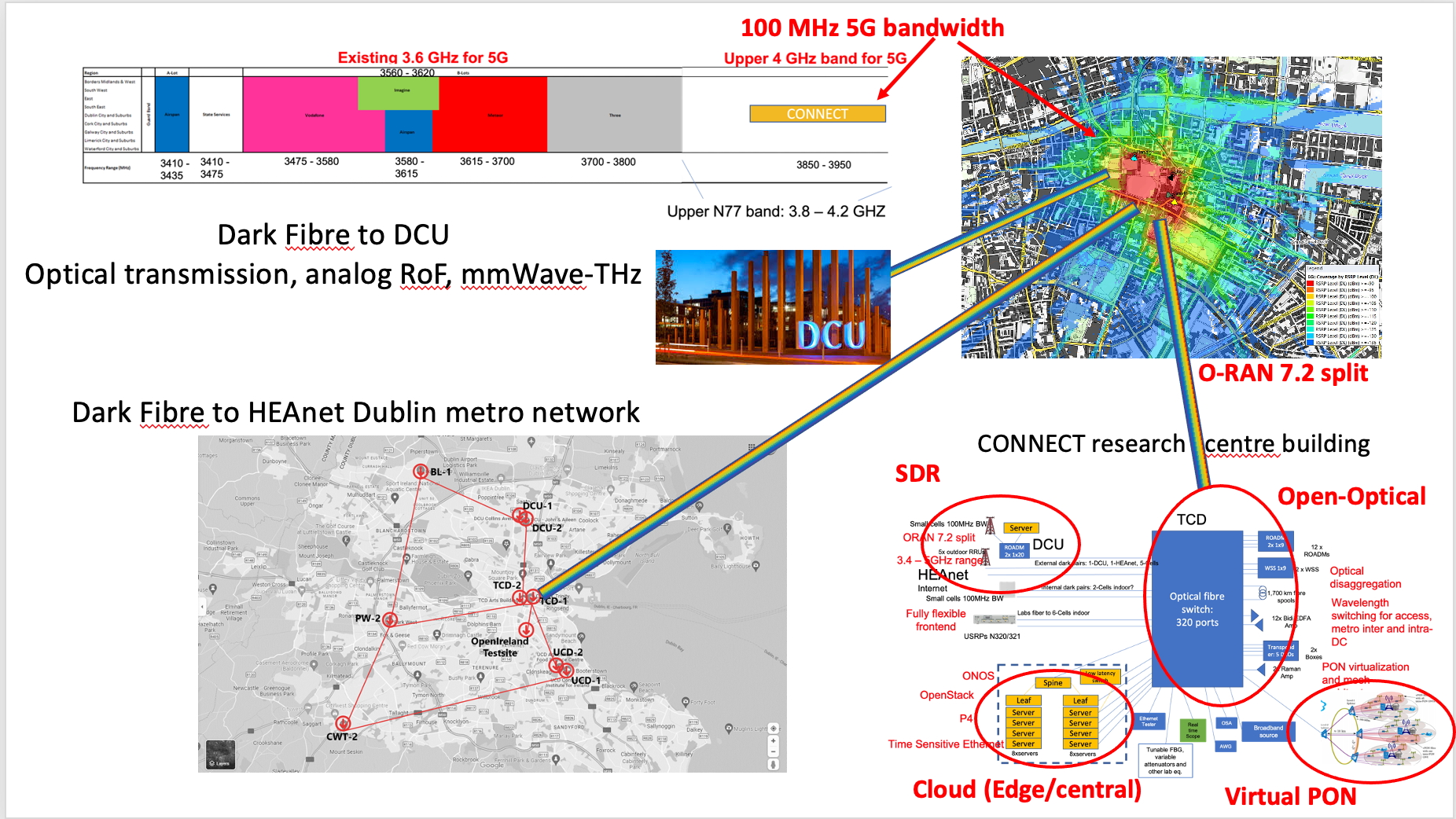}
    \caption{Overall architecture of the OpenIreland testbed.}
    \label{fig:open_ireland_architecture}
\end{figure*}



The new ecosystem that is powered by the flexible fronthaul functional splits, open interfaces and intelligence in all parts of the network enables the long promised open business market in the telecommunication industry. We can finally move on from simple infrastructure sharing to whole business models being based on infrastructure providers, resource providers/brokers, network control and optimization providers and service providers. In such an ecosystem, each entity can focus on optimizing its own part of the end-to-end system. Each of them can also be creative in the interaction with the other stakeholders to stay competitive with price and quality of service. 

Service providers can be as large as existing national telecom operators, but also very small (e.g. a small regional/local or private service provider). Resource providers/brokers can own resources (e.g. wireless spectrum, computing resources) and offer it to service providers or other brokers for a limited time period and a specific geographical location. Pricing can be set depending on the location and the duration of use, which would enable a "stock market" like environment to buy, sell and lease resources depending on the use-case. Network control and optimization providers can focus on developing intelligent mechanisms that can extract information from the open networks and use these insights to dynamically optimize the network for the desired use-cases. These mechanisms can be offered to service providers to either optimize the end-to-end network operation or to provide insights in specific aspects of the network behavior that can further be used by the service providers to optimize their own control mechanisms. Finally, service providers can focus on the end-to-end network operation and its optimization for specific use-cases. 

This approach relaxes the existing "optimize for all user types" problem that current service providers are facing, and will allow small/niche service providers to become feasible/profitable. For example, a service provider that provides \ac{iot} connectivity in remote areas could reduce its cost by leasing infrastructure from different infrastructure providers, i.e. in different remote areas, and lease spectrum only during short periods of times during the day (when the sensors transmit information). This approach reduces cost and removes the need for infrastructure deployment and spectrum license for the time when the spectrum is not actually needed. Similarly, a \ac{v2x} service provider would not require nation-wide resources at all times. To reduce cost, resources could be dynamically acquired only for the locations in which vehicles move. 

The open business market is not a new concept and it has been discussed for a long time \cite{AKYILDIZ201617}. However, there was no push from the telecomms community to open its interfaces end embrace an environment in which large and small businesses compete against each other in a fair strategy game. The \ac{ott} service providers slowly changed the game and operators started losing revenue and became data pipelines \cite{farooq2019impact}. This accelerated the process of opening the industry to new ideas that would allow it to grow into niche areas (e.g. vehicular networks, \ac{uav} networks, \ac{iot}). Nevertheless, flexibility and openness bring many new challenges in terms of technology development, performance optimization, interoperability, control and stability of the network. Therefore, it is necessary to build testbed-like infrastructure projects that allow large scale testing and integration of multi-vendor equipment, different \ac{ran} functional splits and a variety of different node placements. As an example of a testbed that offers functionalities that enable future end-to-end network research, we will provide a detailed description of our OpenIreland testbed. 

\subsection{Problem Description}
Considering the rich business ecosystem described above, a wide range of problems can be addressed, e.g. radio resource sharing, infrastructure sharing, coverage optimization, mobility prediction, power consumption optimization and many others. 

\textit{The problem addressed by our research focuses on handover prediction to minimize the network operation cost through dynamic resource allocation. The problem can be formulated as an optimization problem with equation~\eqref{eq:opt_problem}.}
\begin{equation}\label{eq:opt_problem}
    \begin{aligned}
\text{minimize} &\sum\limits_{t}c_{p_t}\cdot p_t\cdot (1-r_t) + c_{n_t}\cdot r_t\cdot (1-p_t)\\
\textrm{s.t.} \quad & p_t \in \{0,1\} \quad, \quad \forall{t} \in T\\
    &r_t  \in \{0,1\} \quad, \quad \forall{t} \in T\\
    &0 \leq c_{n_t} \leq 1 \quad,\quad\forall{t} \in T\\
    &0 \leq c_{p_t} \leq 1 \quad,\quad\forall{t} \in T\\
\end{aligned}
\end{equation}

The cost function in equation~\eqref{eq:opt_problem} consists of two distinct costs: $C_p$ - the cost of paying for resources when they are not required by the \ac{ue}, which can be defined as: 

\begin{equation}\label{eq:cost_of_paying}
    C_p = \sum\limits_{t}c_{p_t}\cdot p_t\cdot (1-r_t)
\end{equation}
, and $C_n$ - the cost of not paying for resources when they are required by the \ac{ue}, which can be expressed as: 
\begin{equation}\label{eq:cost_of_not_paying}
    C_n = \sum\limits_{t}c_{n_t}\cdot r_t\cdot (1-p_t)
\end{equation}

$r_t$ is a binary variable that indicates whether resources are required in time step $t$. $c_{p_t}$ and $c_{n_t}$ represent the normalized cost of paying and not paying for a resource in the cases when the resource is indeed not required and when it is required by the \ac{ue} in time step $t$, respectively. Finally, $p_t$ is a binary variable that indicates whether the resource was purchased in time step $t$. 

As highlighted earlier, depending on the business model, different stakeholders define different costs, i.e. $c_{p_t}$ and $c_{n_t}$. The services providers can optimize the cost defined by either equation~\eqref{eq:cost_of_paying} or \eqref{eq:cost_of_not_paying}. For example, for an \ac{iot} provider that collects temperature measurements from remote locations the cost $c_{n_t}$ is low due to the fact that the information collected can be buffered and doesn't have a strict latency requirement. Therefore, such a provider would optimize for $C_p$. 

However, businesses with many mobile users will require highly reliable and constantly available access to the network. Therefore, a \ac{v2x} provider would define $c_{n_t}$ as high due to the fact that all information transmitted through the network have strict latency requirements and the failure to provide service at any point of time has a great impact on the provided service. Therefore, such providers would optimize for $C_n$.

 
In equation~\eqref{eq:opt_problem} $r_t$ depends on the user behavior (e.g. user mobility or handover events occurring). In other words, it is not a variable that can be controlled. Similarly, as previously highlighted, the $c_{n_t}$ and $c_{p_t}$ are defined according to the application objective. Therefore, $p_t$ is the only variable that can be controlled by an intelligent algorithm, and it corresponds to the decision made to acquire a resource in time step $t$. This decision was traditionally made by defining coverage areas and buying resources for longer periods of time (e.g. days, months or longer). However, the flexibility introduced by O-RAN allows us to make those decisions at finer granularity. 

Therefore, our goal is to design a \ac{ml} algorithm that predicts handover events, which will be used for decision making in acquiring resources in a given coverage area, with the goal to minimize the cost defined in equation~\eqref{eq:opt_problem}. In other words, the decision to acquire resources $p_t$ in time step $t$ directly relates to the handover prediction made by the \ac{ml} algorithm. If a handover is predicted, that indicates that resources will be required in the new coverage area and therefore $p_t = 1$. On the other hand, if a handover is not predicted, the assumption is that resources will not be required and therefore $p_t = 0$. 

Section~\ref{xapp_provider} will provide more details on how to design an algorithm that considers near-\ac{rt} \ac{ran} data in the decision making process and how to optimize it for $C_p$ and $C_n$.

\subsection{Testbed facilities}
OpenIreland (\url{http://www.openireland.eu}) is state-funded multi-million research infrastructure, which, as shown in Figure~\ref{fig:open_ireland_architecture}, brings together O-RAN infrastructure (indoor and outdoor, open-source and commercial implementations), optical transmission infrastructure (access and metro network) and computing infrastructure. It is open both because it supports new open interfaces for the network control, and because it’s accessible by researchers outside our research group, in industry and academia.

The testbed includes technology that spans optical transmission in access and metropolitan areas (several \acp{roadm} and over $1,700$ km of fiber), 5G new radio systems, both indoor and outdoor and cloud computing technology. It is based on open source software (\emph{Openstack}, \emph{Open-Source MANO}, \emph{Open Network Operating System}, \emph{Goldstone}, \emph{OpenAirInterface} and \emph{srsRAN}) and implemented over open interfaces (i.e., \emph{O-RAN}, \emph{Open and Disaggregated Transport Network}), although it can also provide proprietary solutions. All wireless, optical network and computing elements are connected through a reconfigurable optical switch, allowing us to quickly create different network topologies and scenarios, to study for example the effect that different architectures,latency, data rates and impairments have on the operation of different functional splits in the \ac{ran}.

One of the key aspects is that the testbed enables research and experimentation across three domains (i.e. access, transport and core), as well as the convergence of optical, wireless and cloud computing. In addition it can make real data available to researchers, which is a key factor for the development and testing of new \ac{ml} algorithms. It also links to other universities in Ireland, such as Dublin College University (dark fiber) and abroad, for example the RARE European testbed and the COSMOS testbed in the US (10G data links). 

\begin{figure*}
    \centering
    \includegraphics[scale=0.55]{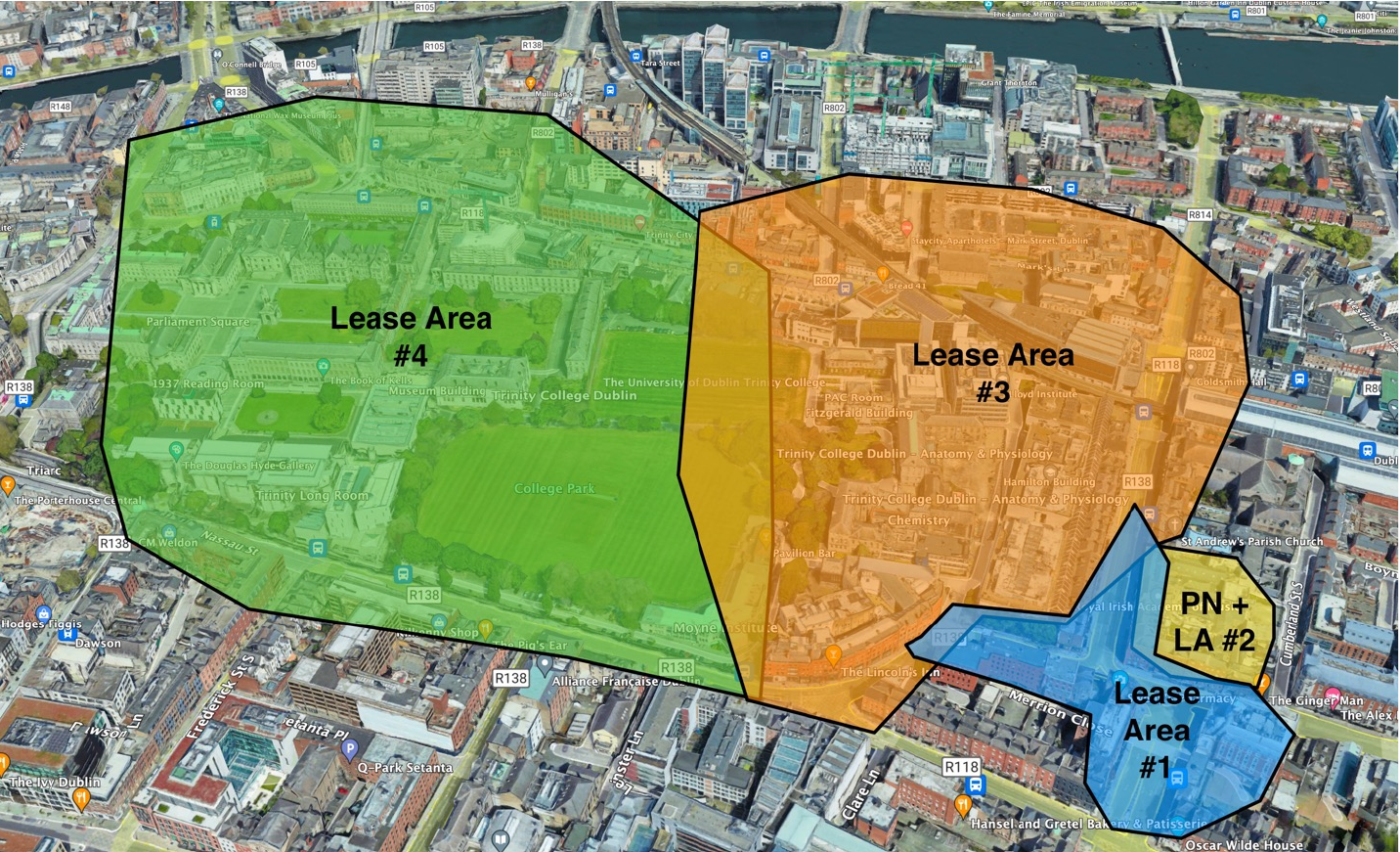}
    \caption{Resource allocation areas, located around the Trinity College Dublin campus.}
    \label{fig:rivada_spectrum_allocation}
\end{figure*}



OpenIreland also comes with two key and unique advantage points (see Figure~\ref{fig:open_ireland_architecture}):
\begin{enumerate}
    \item It enables to carry out research using commercial radio systems (we are in possession of a 100MHz 5G band test license from the Irish telecomms regulator, ComReg) and end-to-end networking (i.e., including optical transmission to include fronthaul/backhaul connectivity). 
    \item It can provide different types of connectivity (from direct fibre, to VPN over the Internet) to a metro public cloud in Dublin, where the cloud experimentation is carried out. This enables exploration of the effect of different latency and jitter experience through different realistic fronthaul and backhaul connectivity modes. 
\end{enumerate}

\subsection{Experimental setup}
In the OpenIreland testbed described in the previous section, we have a working commercial setup of the $7.2$ split with $1$ \ac{cu}, $4$ \acp{du} and $4$ \acp{ru}. On top of the \ac{ran}, our testbed hosts a commercial core network implementation. This setup also includes a commercial 5G spectrum license for $100$MHz of spectrum between $3850$MHz - $3950$MHz. The $4$ \acp{ru} are deployed in and near the Trinity College Dublin campus, situated in Dublin's city center, Ireland. Figure~\ref{fig:rivada_spectrum_allocation} depicts the coverage areas of all $4$ \acp{ru}. Our setup also includes a commercial near-\ac{rt} \ac{ric}. All functional elements were provided by different vendors, allowing us to test the interoperability and open interfaces. 

The commercial setup is used to demonstrate the interoperability between different infrastructure providers, but also to understand the role resource brokers and network control and optimization providers can have on the service design carried out by service providers. More precisely, we rely on a spectrum sharing approach proposed by one of our commercial partners (Rivada Networks) that allows service providers to buy/lease spectrum in different regions and assign their users to the allocated spectrum resources. 

Figure~\ref{fig:rivada_spectrum_allocation} shows $4$ different resource allocation areas. Each area has spectrum allocated to it and a service provider can lease that spectrum for its own users. The same spectrum can (but does not have to) be shared between different service providers depending on the user allocation. A service provider can dynamically allocate its users to different coverage areas or restrict users from accessing parts of the spectrum. This allows the service provider to only pay for the spectrum when it is needed and to reduce cost by providing access only to users that require coverage in specific geographical regions. This setup allows us to demonstrate the full business potential of O-RAN as discussed in Section~\ref{sec:problem_description_and_experimental_setup}.

The above-mentioned approach allows small service providers that focus on a niche technology to enter the market and reduce cost by intelligently requesting infrastructure and spectrum resources. However, in order to optimize its operation, a control plane is required to collect information from the network and to dynamically reconfigure the network resources. As previously highlighted, we will focus on the mobility use-case. In other words, we are envisioning a service provider that offers connectivity to users that move through the network. However, instead of deploying its own infrastructure and buying expensive annual spectrum licenses, the service provider will collect information from the network to make predictions about location-based resource requirements and reduce cost by paying only for resources it predicts the users will need. Instead of opting for an extreme use-case, focusing on either the optimization of $C_p$ or $C_n$, our approach will be more generic and discuss both options as well as hybrid approaches. 

The O-RAN architecture with its various control planes and open interfaces offers the required network programmability and data collection capability to build intelligent algorithms that can predict resource needs before they arise. Considering the described mobility use-case, we will propose a \ac{ml} approach to predict handovers based on the data collected from the \ac{ran} over the open interfaces within our OpenIreland testbed. The \ac{ml} predictions provide important insights to the service providers for decision making tasks related to leasing resources.


\section{xApp Design}\label{xapp_provider}

An xApp is an extended application that runs on the near-RT RIC. It relies on \acp{sm} to collect performance metrics and sends control messages to the \ac{ran}. \acp{sm} can be viewed as interfaces that are used for the communication between E2 endpoints. Our xApp runs on a commercial setup available in our testbed and relies on the E2SM-KPM \cite{e2sm-kpm} \ac{sm} for data collection. Besides the availability of \acp{sm}, there are multiple steps involved in the process of building an xApp that uses \ac{ml} to provide valuable information to network stakeholders: (1) Data collection (E2SM-KPM); (2) \ac{ml} workflow; and (3) Stakeholder adaptation. These steps have to be mapped on the correct nodes in the architecture and incorporated in the \ac{ml} workflow described in Section~\ref{sec:introduction}.

\subsection{Data Collection}\label{sec:data_collection}
As shown in Figure~\ref{fig:rivada_spectrum_allocation}, our network consists of $4$ \acp{bs} (i.e. $1$ indoor and $3$ outdoor). More precisely, the network architecture includes $4$ \acp{ru}, $4$ \acp{du} ($1$ per \acp{ru}) and $1$ \ac{cu}. All \acp{ru} have a $40$MHz band of spectrum assigned to them from the $100$MHz available to our OpenIreland testbed. The network architecture also includes a near-\ac{rt} \ac{ric}, which hosts our xApp that was used for data collection. As previously shown in Figure~\ref{fig:oran_architecture}, our xApp communicates with the \ac{du} and \ac{cu} through the \emph{E2} interface. 

In general, the handover process is divided into three steps: measurement, judgment, and execution. The measurement process is the crucial step, since it informs the other two. For the purpose of making a handover decision, the \ac{rsrp} and \ac{rsrq} are the most important measurements. The judgment process considers these two measurements and compares the serving cell measurements to the neighboring cells. In case the serving cell measurements are lower than the neighboring cell measurements by a defined margin, the handover process will be triggered. 

The xApp has access to various measurements from the \ac{ran}, ranging from node status updates to \ac{ue} specific measurements. For the purpose of our experiments, we are interested in the \ac{ue} measurements related to signal and interference levels. More precisely, we are interested in the measurements of the \ac{rsrp}, \ac{rsrq} and \ac{sinr} for the serving cell and all neighboring cells within range. The \ac{rsrp} is the average received power of \acp{re} that carry cell specific \acp{rs} over the entire bandwidth. The \ac{rsrq} is defined as the ratio of the carrier power to the interference power. In other words, it is a signal to noise ratio measured using a standard signal. Equation~\eqref{eq:rsrq} shows the relationship between the \ac{rsrq}, \ac{rsrp} and \ac{rssi}. 

\begin{equation}\label{eq:rsrq}
    RSRQ = \frac{N \cdot RSRP}{RSSI},
\end{equation}
where $N$ is the number of \acp{rb} per channel bandwidth.

The \ac{rsrp} is reported in the range from $-140$dBm to $–44$dBm with $1$dB resolution, and the \ac{rsrq} from $-3$dBm to $-19.5$dBm with $0.5$dB resolution. A minimum of $-20$dB \ac{sinr} (of the \emph{S-Synch channel}) is needed to detect the \ac{rsrp}/\ac{rsrq}. 


Our dataset consists of $40,000$ samples, each containing a timestamp, \ac{rsrp}, \ac{rsrq}, \ac{sinr} measurements for the serving cell and $3$ neighboring cells and indications whether a handover happened or not. Out of all these samples only $4,350$ contain handovers, which is approximately $10\%$ of the total dataset. It is important to notice that this will have an impact on the \ac{ml} model design and training decisions. The time granularity is $1$s, meaning that every second we would collect one sample. Therefore, our dataset contains $11$ hours worth of data. 



\subsection{ML workflow}\label{sec:ml_modeling}

Considering that the handover procedures are well defined, we would expect that a simple threshold would be enough to predict if a handover is about to happen. This is true only if we have a simple model with two \acp{bs} and a single \ac{ue} moving in a straight line between them. Once the system is more complex and obstacles are added to the environment, predicting the handovers involves extracting user mobility patterns from a limited set of features (i.e. \ac{rsrp}, \ac{rsrq} and \ac{sinr}) structured in a sequential dataset (i.e. the dataset that was collected in Section~\ref{sec:data_collection}). 

\begin{table}[t]
\centering
\caption{LSTM model details.}
\label{tab:lstm_model}
\begin{tabular}{l|cccccc}
Layer               & \textbf{Type}             & \textbf{Size}           & \textbf{Dropout} & \textbf{Recurrent Dropout} & \textbf{} & \textbf{} \\ \cline{1-5}
\textbf{Input}      & NA                        & 12 & NA               & NA                         &           &           \\
\textbf{1st hidden} & LSTM & 32                      & 10\%             & 50\%                       &           &           \\
\textbf{2nd hidden} & LSTM                      & 64                      & 10\%             & 50\%                       &           &           \\
\textbf{3rd hidden} & LSTM                      & 32                      & 10\%             & 50\%                       &           &           \\
\textbf{Output}     & Softmax                   & 2                       & NA               & NA 
\end{tabular}
\end{table}

A sequential learning model that takes time distributed data as inputs and predicts handover events fits the problem description very well. However, we have only a limited set of features describing the signal quality and we have no information about the environment in which the \acp{ue} move. This makes the prediction challenging. However, we extract the correlation between short and long term historical measurements to predict the likelihood of handover events happening. For example, if the serving cell signal strength starts going down and then it increases for a short period of time, this might be an indication of the \ac{ue} approaching an obstacle that faces the neighboring cells. While the user was far away from the obstacle, it was within line of sight of the neighboring cell, but once it got closer, shadowing from the obstacle affected its signal strength, resulting in the \ac{ue} remaining connected to the existing serving cell. This is only one example showing that it is not enough to look at thresholds being approached by a \ac{ue} to make predictions about handovers. 

\acp{rnn} are the obvious choice for the problem described above. However, due to the requirement to correlate long and short term memory, \acp{lstm} are a better choice. Additionally, \acp{lstm} remove the issues related to vanishing and exploding gradients. As shown in Table~\ref{tab:lstm_model}, our model consists of an input layer, $3$ hidden \ac{lstm} layers, and a \emph{softmax} output layer. All hidden layers are trained with $10\%$ dropouts and $50\%$ recurrent dropouts for regularization purposes. 

Table~\ref{tab:lstm_model} also shows that the output layer is a \emph{softmax} layer with $2$ nodes. Therefore, our model is a sequence-to-vector model. In other words, the model takes a sequence of measurements and predicts one value for the future. This can be defined as follows:

\begin{equation}\label{eq:model_io}
    \begin{bmatrix} f_1 \\ \vdots \\ f_k \end{bmatrix} \;
    \xrightarrow{}
    Model
    \xrightarrow{}
    \begin{bmatrix} P_1 \\ P_2 \end{bmatrix}, \;
\end{equation}
where $f_1, \ldots, f_k$ are feature vectors sorted chronologically, and $P_1$ and $P_2$ are the two possible predictions, i.e. handover will or will not occur in the next $t$ time steps. A time step in our dataset is equal to $1$s, meaning that we are trying to predict whether a handover will happen in the next $t$ number of seconds. The prediction horizon (i.e. how many seconds we look into the future) is a parameter that depends on the performance of our prediction and the specific requirements of the service level. Both, $k$ and $t$ (i.e. the history and horizon parameters) will be further discussed in Section~\ref{sec:evaluation}.

In case of a correct prediction, both $C_p = 0$ and $C_n = 0$. However, if a handover is predicted in time step $t$, but it does not happen in reality, i.e. $p_t = 1$ and $r_t = 0$, then:
\begin{equation}\label{eq:sum_of_false_positives}
    \begin{aligned}
    C_p &= \sum\limits_{t, p_t=1, r_t = 0}c_{p_t} \\
    C_n &= 0 \quad\\
\end{aligned}
\end{equation}
Considering that the prediction of handovers in this case is a binary classification problem defined by equation~\eqref{eq:model_io}, equation~\eqref{eq:sum_of_false_positives} clearly shows that the overall cost will be equal to the sum of individual $c_{p_t}$ for every $t$ in which our prediction results in a \ac{fp}. Therefore, for a constant $c_{p_t}$, equation~\eqref{eq:sum_of_false_positives} can be written as: $C_p = FP \cdot c_{p}$.

Similarly, if a handover is not predicted for the next $t$ time steps, but it happens in reality, i.e. $p_t = 0$ and $r_t = 1$, then:
\begin{equation}\label{eq:sum_of_false_negatives}
    \begin{aligned}
    C_p &= 0  \quad\\
    C_n &= \sum\limits_{t, p_t=0, r_t = 1}c_{n_t}\\
\end{aligned}
\end{equation}
Again, for a binary classification problem defined by equation~\eqref{eq:model_io}, the cost will correspond to the sum of individual $c_{n_t}$ for every $t$ in which our prediction results in a \ac{fn}. Therefore, the cost in equation~\eqref{eq:sum_of_false_negatives} for a constant $c_{n_t}$ can be written as: $C_n = FN \cdot c_{n}$.

Whenever a \ac{ml} problem is posed as a classification problem, it is important to analyze its performance in terms of precision, recall and F1-score. Additionally, equations~\eqref{eq:sum_of_false_positives} and \eqref{eq:sum_of_false_negatives} indicate that depending on the application level cost optimization choice, the \ac{ml} model should be optimized for either recall or precision in order to optimize the model for either \acp{fp} or \acp{fn}. Recall ($R$) and precision ($P$) are defined in terms of \ac{tp}, \ac{fn}, and \ac{fp} predictions as follows:
\begin{equation}\label{eq:recall_and_precision}
    \begin{aligned}
    R &= \frac{TP}{TP + FN}  \quad\\
    P &= \frac{TP}{TP + FP}\\
\end{aligned}
\end{equation}
The F1-score is defined with: 
\begin{equation}\label{eq:f1-score}
    F1 = 2 \cdot \frac{P \cdot R}{P + R} = \frac{2 \cdot TP}{2 \cdot TP + FP + FN}
\end{equation}

The relationship between recall and precision for the purpose of these kinds of predictions depends on the application level objective. For example, a service provider can choose to optimize the network configuration overall. In other words, they choose similar costs $c_{p_t}$ and $c_{n_t}$ (e.g. both being equal to $0.5$). This means that instead of optimizing for recall or precision, they want to optimize for both (i.e. F1-score), resulting in minimal values for both, recall and precision, combined. According to equation~\eqref{eq:f1-score}, this results in minimal combined \ac{fp} and \ac{fn} and accordingly minimal cost as defined in equation~\eqref{eq:opt_problem}.

On the other hand, a different type of service provider that focuses on non-critical mobility use-cases that are service quality oriented would focus on recall while keeping the precision at a predefined level. The reason is that even though high recall and lower precision would result in resources being leased and paid for and sometimes not being used, this increases the probability that resources are available when needed, thus delivering a higher quality of experience. 
Our evaluation in Section~\ref{sec:evaluation} will focus on this use-case. 

Please note that for the problem at hand, mission critical use-cases, \textit{for which the overall cost is not an issue}, are easier to optimize for by simply reaching recall of $100\%$ (e.g. leasing resources in all areas surrounding the current coverage area).



In Section~\ref{sec:data_collection}, we highlight that the dataset is not balanced, favoring the no handover class. Therefore, besides the common \ac{ml} issues related to the trade-off between bias and high variance, due to the fact that only $10\%$ of the samples represent handover events, we have to consider the class imbalance while training the model. For that purpose, we compute the class weight for both handover ("HO") and no handover ("No-HO") classes. The imbalance and therefore the weights depend on the horizon parameter. The longer the horizon the higher the probability of seeing a handover. 

\begin{equation}
    w_l =  \frac{n}{L \cdot l_n},
\end{equation}
where $w_l$ is the weight of class $l$, $n$ is the overall number of samples, $L$ is the number of classes and $l_n$ is the number of samples that belong to class $l$. This allows us to assign higher weights to the minority class, which helps the model to pay more attention to its patterns and reduce bias towards the majority class.

To summarize, the decision of which \ac{ml} model to use for enforcing network policies is a high-level strategic decision made by the service provider, while the training of the models and the inference over network traffic takes place in the non-RT and near-RT \acp{ric}, respectively. This separation of responsibilities allows for a clear distinction between the strategic decision-making process and the operational execution of network policies (see Figure~\ref{fig:oran_architecture}):
\begin{enumerate}
    \item (non-RT RIC): At a strategic level, factors such as the desired network performance and cost constraints are considered when selecting the appropriate \ac{ml} model. Once a model is chosen, it is trained in the non-RT RIC, which has sufficient resources and processing power to handle the complex training process.
    \item (near-RT RIC): After training, the \ac{ml} model is submitted to the near-RT \ac{ric}, which is responsible for enforcing network policies in near real time. The near-RT \ac{ric} analyzes incoming traffic and applies the \ac{ml} model to make decisions about resource acquisition in neighboring coverage areas. 
\end{enumerate}


\subsection{Stakeholder adaptation}
\begin{figure}[t]
    \centering
    \includegraphics[scale=0.35]{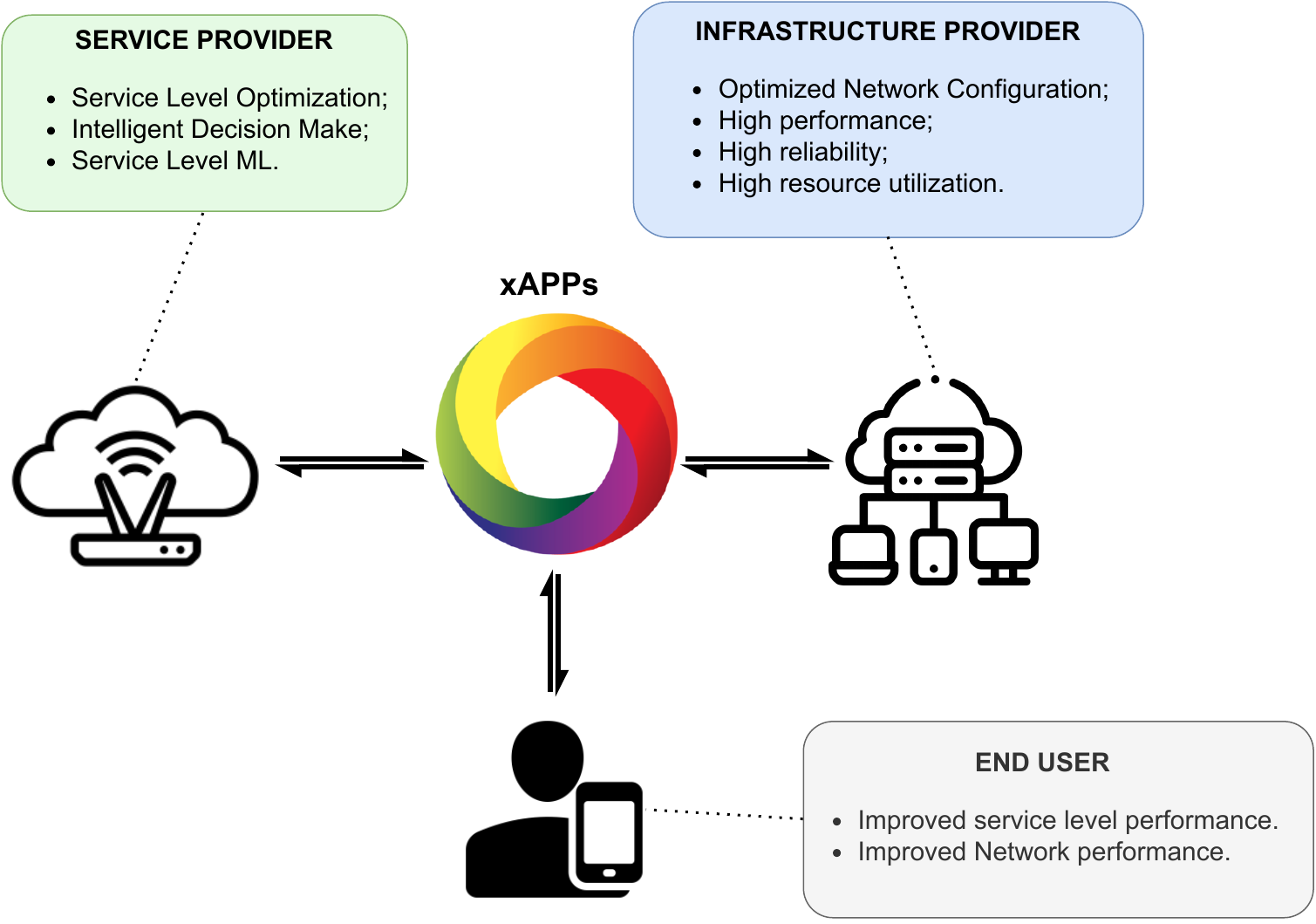}
    \caption{The benefits various stakeholders gain from the use of information provided by an O-RAN intelligent control mechanism (i.e. xApp).}
    \label{fig:network_stakeholders}
    \vspace{-1.5em}
\end{figure}

An xApp as described in the previous section provides valuable information to all stakeholders that rely on communication networks to conduct their business. As shown in Figure~\ref{fig:network_stakeholders}, users benefit from an improved network and service level experience, while infrastructure providers benefit from an optimized network configuration resulting in high performance, reliability, and resource utilization. Lastly, service providers benefit from the information for the purpose of service optimization, on-demand resource leasing decisions and by feeding the information from various xApps to their own service level learning algorithms.

This opens a new marketplace for network control and optimization providers. The fact that the open interfaces allow them to integrate with different vendors, and that the information they provide are valuable to multiple entities allow them to offer their services at scale. Unlike the traditional approach to network optimization, where expert knowledge has to be applied to the equipment of different vendors, and datasets have to be exported and analyzed before reconfiguration decision can be made, the new approach that relies on \ac{ml} and open interfaces adapts to the environment it operates in. 

Depending on the use-case, the xApps can be implemented to: (1) extract and pre-process information from the network; (2) make valuable predictions about network processes (e.g. user behavior, traffic load, traffic classes); and (3) reconfigure different network nodes. However, an xApp does not have to implement all three steps to provide value. For example, an xApp that just extracts information from the network and presents it in a meaningful way can be used as a replacement for the majority of the network monitoring tools (e.g. alarm monitoring, \ac{kpi} monitoring). The extracted information can also be fed to other learning algorithms (e.g. other xApps or rApps). 
Once the loop is closed with the reconfiguration of different network nodes, the xApp can be used as a standalone entity that monitors and improves the network/service level performance. 

In line with the above-mentioned xApp implementation types, the handover prediction proposed in this paper implements the first and second steps, i.e. information extraction and pre-processing, and \ac{ml}-based handover predictions that can be shared with all interested stakeholders. The predictions are further used to make resource acquisition decisions, which similarly to our approach can further be optimized to acquire resource for the long/short term, for a larger/smaller area, from one/multiple vendors. 

\begin{figure}
    \centering
    \includegraphics[scale=0.5]{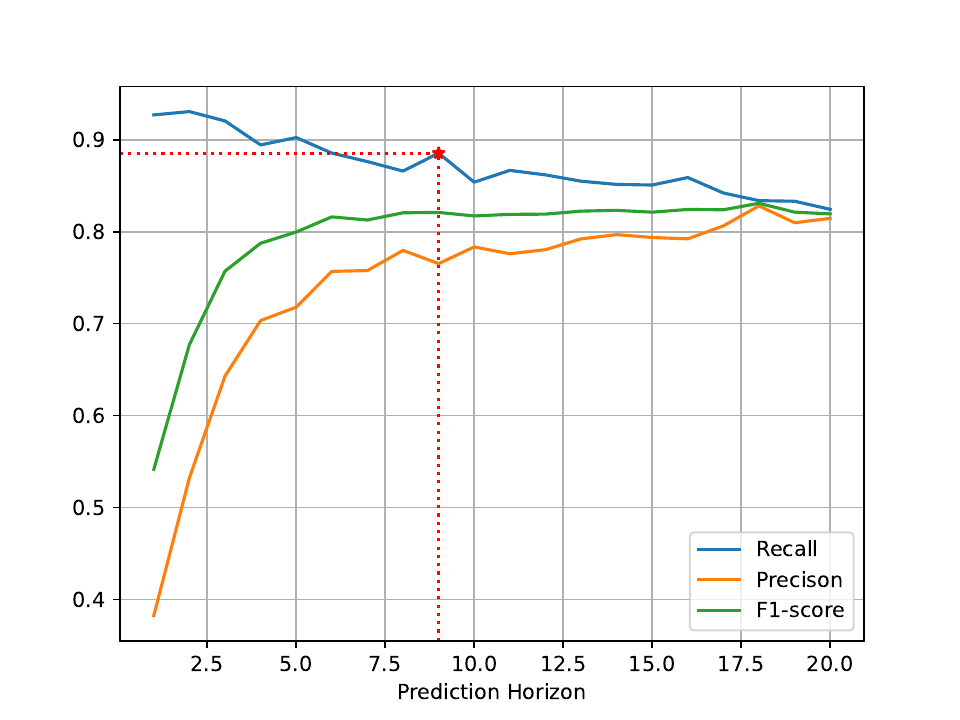}
    \caption{Recall, precision and F1-score depending on the prediction horizon. The prediction history is fixed at $10s$.}
    \label{fig:recall_precision_f1_score}
    \vspace{-1.5em}
\end{figure}
\section{ML Model Performance Evaluation}\label{sec:evaluation}
In this section we will evaluate the performance of our models and explain the reasoning behind the chosen features, the prediction history and horizon as well as the importance of understanding the value of recall vs precision for the purpose of predicting handovers depending on the xApp user.


As previously highlighted in Section~\ref{xapp_provider}, in addition to the handover margin defined by the service provider, the \ac{rsrp}, \ac{rsrq} and \ac{sinr} are the most important measurements that determine when a handover should happen. In Section~\ref{xapp_provider} we also explain that when operating in a real environment, the handover prediction is not as straight forward as learning the threshold and margin values and that it also depends on the environment in which the user moves, obstacles the user faces and the user mobility patterns. 

We take these measurements and feed them to the \ac{lstm} network presented in Section~\ref{sec:ml_modeling}. The measurement and horizon time-step resolution is $1s$, meaning that having a horizon prediction of $10$ time-steps is equivalent to predicting whether a handover will happen in the next $10s$. We use a $60-20-20$ split of the dataset to train, validate and test the model. The results of the prediction are evaluated on the testing dataset. The model performs a classification tasks to distinguish between HO and No-HO in the prediction horizon time frame. 

\begin{figure}
    \centering
    \includegraphics[scale=0.5]{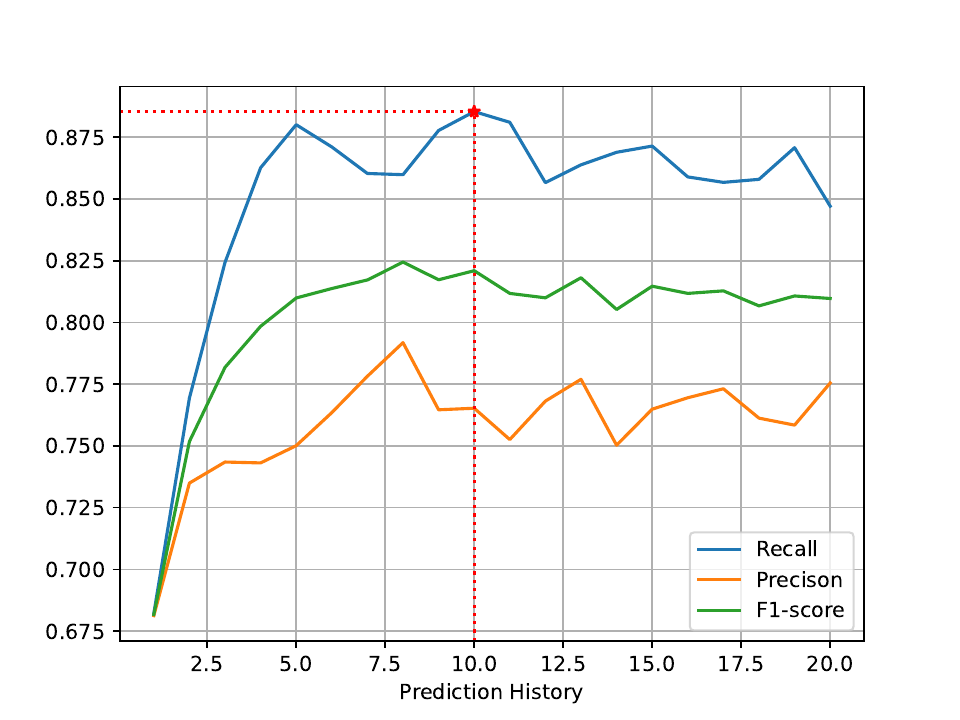}
    \caption{Recall, precision and F1-score depending on the prediction history. The prediction horizon is fixed at $9s$.}
    \label{fig:recall_precision_f1_score_history}
    \vspace{-1em}
\end{figure}

\textcolor{corr1}{Time series forecasting models, such as \acp{lstm}, require the specification of both prediction history and prediction horizon. Prediction history includes the data points that will be fed into the model to generate future predictions. In essence, prediction history determines the number of time steps we will examine in the past. The prediction horizon, on the other hand, specifies how far into the future we are attempting to predict. It refers to the number of time steps in the future for which we will attempt to make a prediction based on the historical data utilized.}

According to 3GPP recommendations, the average coverage radius in the urban environment is $500m$ \cite{3gpp_38_913_v_14_2_r_14}. For the purpose of handover prediction, we are not interested in the whole coverage area. Our main focus is on the $10\%$ of the radius (approx. $50m$) that is close to the border with another cell. Considering that the average walking speed is around $1.4m/s$ and that the average driving speed in the city is around $1.4m/s$ (i.e., $40km/h$), the average time it takes a user moving at approx. $6m/s$ (which is the arithmetic mean between these two extremes) to move across the distance of $50-60m$ is $10s$. Therefore, we first start by fixing the prediction history to $10s$ and analyzing different prediction horizons. 

As highlighted earlier in Section~\ref{sec:ml_modeling}, we examine a use-case which includes optimizing an \ac{ml} model for non-critical mobility services. In other words, our focus is on optimizing the model for recall, with a predefined minimum precision threshold equal to $75\%$. 

Figure~\ref{fig:recall_precision_f1_score} shows how the recall, precision and F1-score change depending on the chosen prediction horizon. \textcolor{corr1}{Our goal is to acquire resources only if we predict that a handover will happen. Figure~\ref{fig:recall_precision_f1_score} shows that by choosing a prediction horizon that is too short, precision will be very low, while recall is very high. This is due to the fact that predicting the exact moment of a handover event is very hard and that in a networking dataset handovers are rare events. Therefore, when the measurements change and suggest that a handover will happen soon, the model starts predicting handovers before they actually happen, which results in a large number of \acp{fp}. By extending the horizon, we are relaxing the constraints of the prediction and instead of predicting the exact moment of the handover event, we are predicting a time range within which a handover will occur. If the prediction horizon is very long, it is easy to predict a handover event. The longer the prediction horizon the higher the probability of a handover event occurring. Therefore, the precision increases with an increasing horizon. However, if the prediction horizon is too long, we are paying the cost of acquiring resources that are not being used. Therefore, the goal is to have a horizon that is long enough to allow us to acquire resources and to improve the precision, and short enough to avoid paying for resources while they are not in use. A closer look at Figure~\ref{fig:recall_precision_f1_score} shows that the horizon of $9s$ results in high precision and recall. Therefore, we choose this prediction horizon for our model.}

Now that the horizon is chosen, we re-examine the history timescale to ensure best prediction performance. Figure~\ref{fig:recall_precision_f1_score_history} provides additional information about how the prediction history affects the performance metrics. The figure depicts the recall, precision and F1-score for a fixed prediction horizon equal to $9s$ and prediction histories ranging from $1s$ to $20s$. Reducing the history down to $5s$ does not affect the performance of our predictions, but further reduction decreases the performance significantly. On the other hand, increasing the prediction history above $10s$ will not result in improved prediction performance. This is due to the fact that "old" measurements are becoming irrelevant for the prediction about the user mobility. The value that maximizes recall is $10s$. 

\begin{figure}
    \centering
    \includegraphics[scale=0.5]{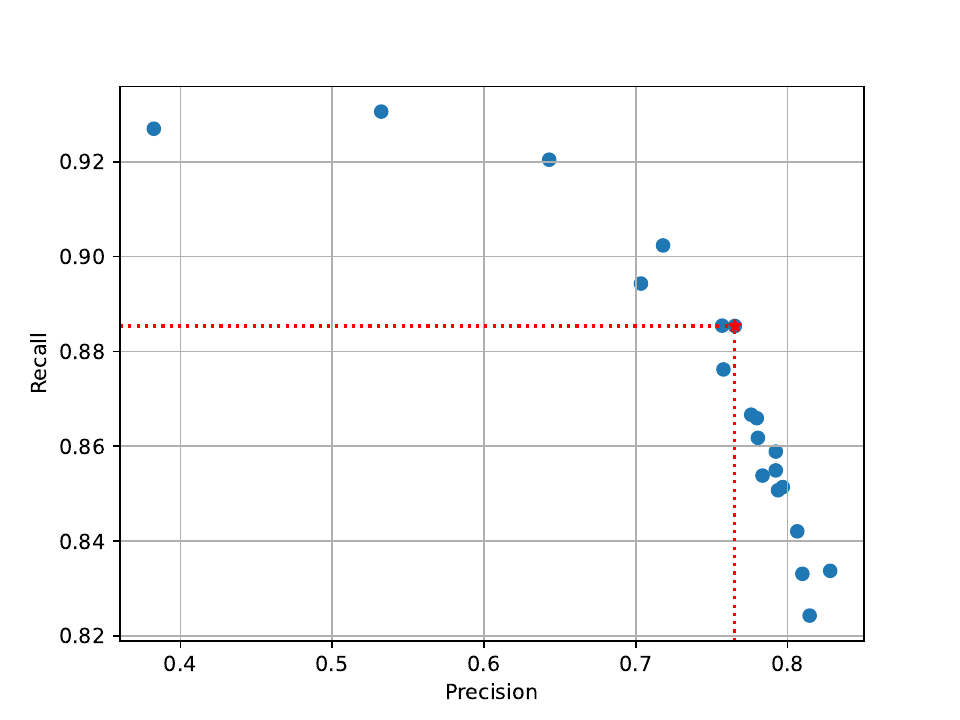}
    \caption{Recall vs Precision}
    \label{fig:recall_vs_precision}
    \vspace{-1.5em}
\end{figure}

\begin{figure}
    \centering
    \includegraphics[scale=0.4]{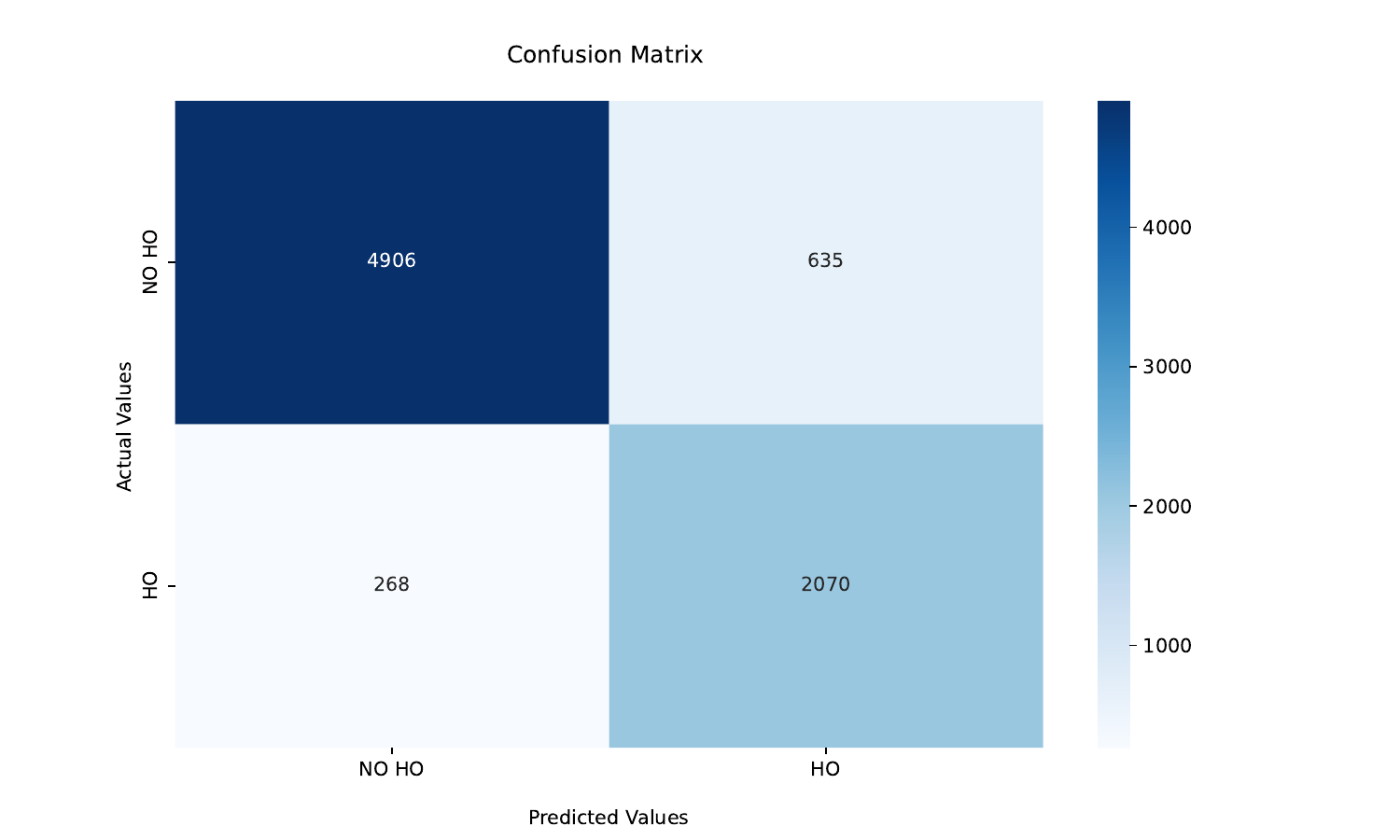}
    \caption{Confusion matrix for the chosen model ($10s$ prediction history and $9s$ prediction horizon).}
    \label{fig:confusion_matrix}
    \vspace{-1em}
\end{figure}

\begin{figure*}
    \centering
    \begin{subfigure}[b]{0.48\textwidth}
    \centering
        \includegraphics[trim=478 40 400 60,clip,scale=0.8]{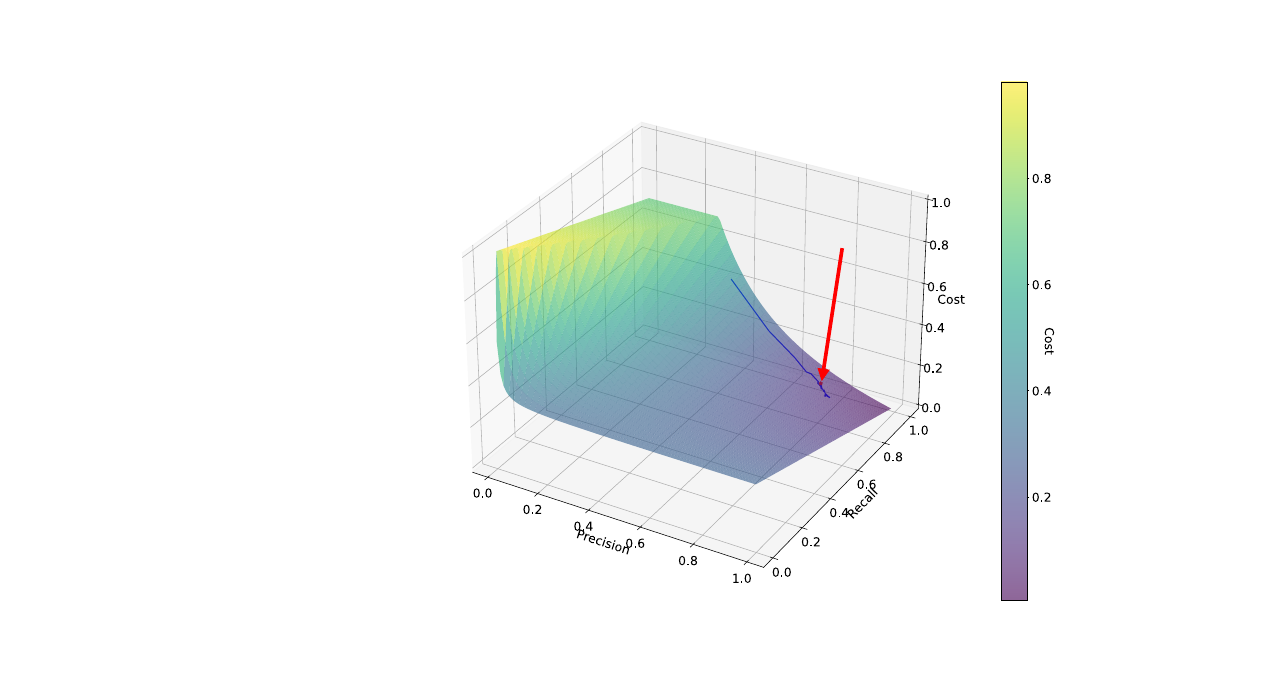}
    \end{subfigure}
    \hfill
    \begin{subfigure}[b]{0.48\textwidth}
    \centering
        \includegraphics[trim=478 40 400 60,clip,scale=0.8]{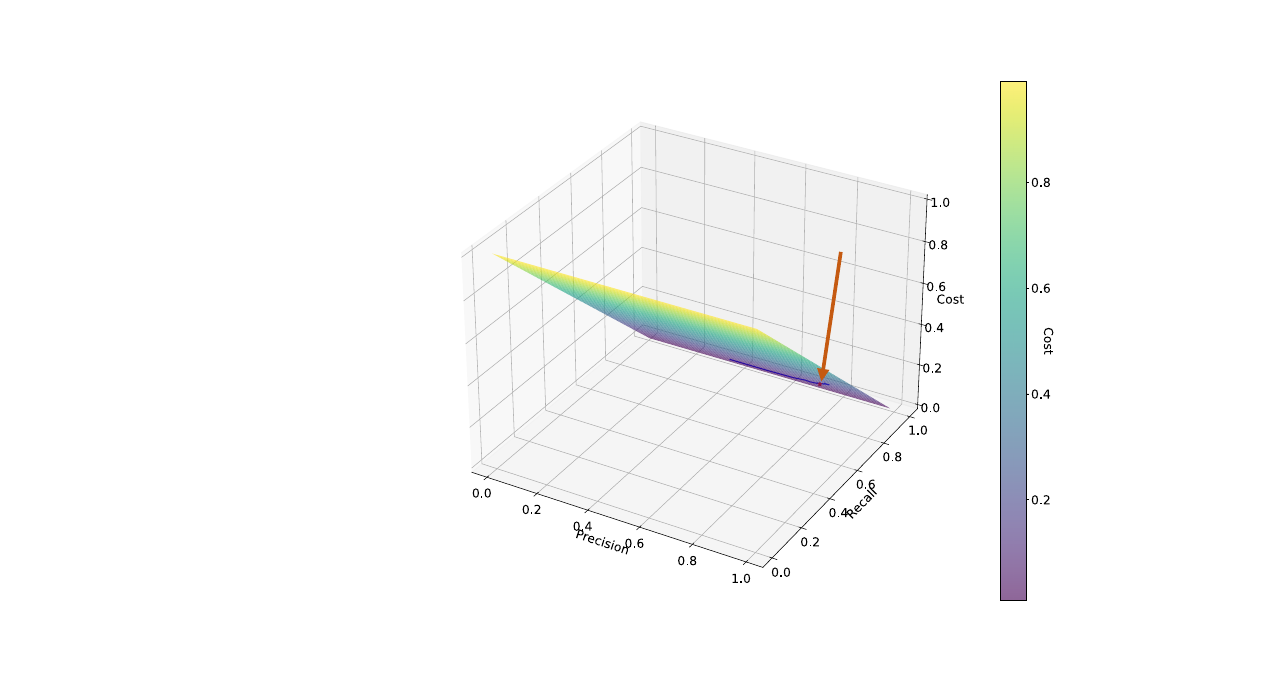}    
    \end{subfigure}
    \hfill
    \begin{subfigure}[b]{0.48\textwidth}
    \centering
        \includegraphics[trim=540 25 400 35,clip,scale=0.8]{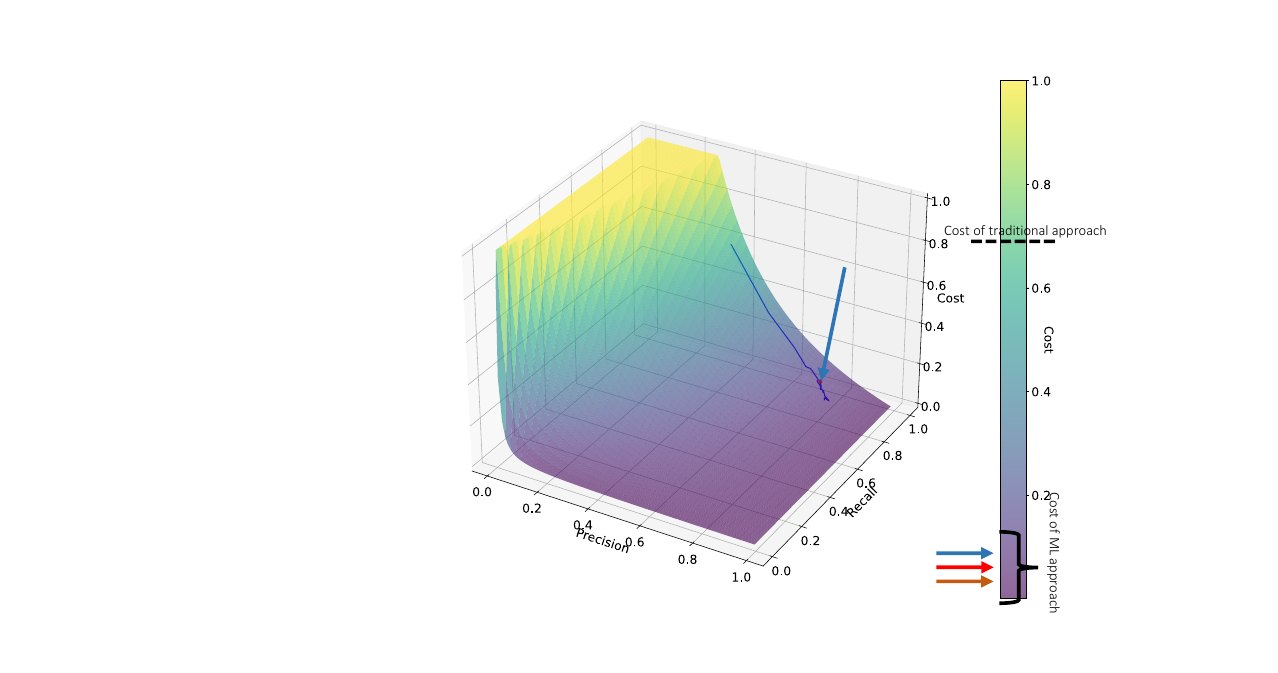}    
    \end{subfigure}
    \caption{Cost depending vs Precision and Recall. From left to right, the configurations are as follows: ($c_{p_t}=1$, $c_{n_t}=1$), ($c_{p_t}=0$, $c_{n_t}=1$), ($c_{p_t}=1$, $c_{n_t}=0$)}
    \label{fig:cp1_cn1}
\end{figure*}

In \ac{ml}, a good approach to choose a good trade-off between precision and recall is to use a precision-vs-recall curve. The precision-vs-recall curve for our models is shown in Figure~\ref{fig:recall_vs_precision}. 
\textcolor{corr1}{The Figure was created by computing recall and precision values of all models trained with different prediction horizons. To choose the best model that optimizes both metrics, we choose a model that corresponds to a point on the right side of the graph, just before the recall values drops significantly. This allows us to choose a model with high recall, that will also result in acceptably high precision.}
The red star indicates the chosen model and its corresponding recall and precision values. Please note that this analysis has been done on the validation set. As shown in Figure~\ref{fig:recall_vs_precision}, the precision value for the chosen recall is very close to $80\%$. 

\textcolor{corr1}{Besides the obvious trade-off between recall and precision, Figure~\ref{fig:recall_vs_precision} indirectly provides information on the trade-off between $C_p$ and $C_n$. Therefore, depending on the relative importance of the individual costs, the \ac{ml} models can be chosen to accommodate the specific need of a service provider.} Since precision directly affects the cost of acquiring resources - $C_p$, according to Figure~\ref{fig:recall_vs_precision} further increase of the recall makes it costly for the service provider due to the significant precision drop. On the other hand, recall affect the cost of resources not being available when they are required - $C_n$. Figure~\ref{fig:recall_vs_precision} suggests that models resulting in higher precision values will significantly increase that cost by reducing recall.


Figure~\ref{fig:confusion_matrix} shows the confusion matrix for the chosen model. The figure clearly shows that the model is able to predict the handovers with high precision and recall, making only a small number of prediction errors. Considering that the model was optimized for recall, the fact that the number of \emph{False Negatives} is smaller compared to the number of \emph{False Positives} is to be expected.

Figure~\ref{fig:cp1_cn1} depicts the network operation cost associated with resource acquisition depending on the service requirements. \textcolor{corr2}{These are based on the concept that service providers outline the cost models based on \acp{sla}, and these models are then translated into cost parameters for equation~\eqref{eq:opt_problem}.} We analyze three alternative use cases: 
\begin{enumerate}
    \item The case of a service provider that wants to compromise between the risk of over-booking resources (i.e., booking when they are not needed) and the risk of lacking resources when they are required by the users. This is translated on setting the cost parameters $c_{p_t}=1$ and $c_{n_t}=1$ \textcolor{corr2}{and this corresponds to the surface plot on the top left of Figure~\ref{fig:cp1_cn1}};
    \item The case of a service provider aiming for high availability, which penalizes the lack of resources but does not penalize over-booking , i.e. $c_{p_t}=0$ and $c_{n_t}=1$ \textcolor{corr2}{and this corresponds to the surface plot on the top right of Figure~\ref{fig:cp1_cn1}};
    \item The case of a service provider aiming for minimize cost of resources (e.g. IoT provider for non real-time services), which penalizes the acquisition of resources when they are not required, but does not penalize lack of resources when they are required, i.e., $c_{p_t}=1$ and $c_{n_t}=0$, \textcolor{corr2}{which corresponds to the surface plot on the bottom of Figure~\ref{fig:cp1_cn1}}.
\end{enumerate}


\textcolor{corr2}{While in the figures we present these three extreme cases, any combination of $c_{p_t}$ and $c_{n_t}$ could be chosen to fine tune the trade-off between availability of resources and cost of network operation. Considering the direct relationship between these costs and precision/recall, \ac{ml} models can be optimized to support any cost model outlined in the \ac{sla}.} The surface plots in Figure~\ref{fig:cp1_cn1} show the possible costs for all viable combinations of recall and precision, based on the cost definition introduced in equation~\eqref{eq:opt_problem}. \textcolor{corr2}{In other words, these surface plots represent the feasible solution space for the chosen \ac{sla}.} Obviously, precision and recall are not completely independent values. For example, equations~\eqref{eq:recall_and_precision} show that both values will be equal to $0$ if the number of \acs{tp} is equal to $0$. The connection over the \acp{tp} will inform the viable precision values for any chosen recall value. The shape of the surfaces depends on the penalty definitions, which might be outlined on \acp{sla}. 

The overall cost of operation as defined in equation~\eqref{eq:opt_problem} will then only depend on the performance of the \ac{ml} model. Therefore, we overlay (blue line) the cost of all models trained in our evaluation on top of the three surface plots shown in Figure~\ref{fig:cp1_cn1}. This line shows the cost associated with all models corresponding to different precision-recall values depicted in Figure~\ref{fig:recall_vs_precision}. The overall cost in Figure~\ref{fig:cp1_cn1} is normalized, resulting in values from [$0-1$]. 
All three plots in Figure~\ref{fig:cp1_cn1} confirm that the overall cost is equal to $0$ when precision and recall are both equal to $1$. Similarly, they all confirm that the overall cost is equal to $1$ when precision and recall are both equal to $0$. \textcolor{corr3}{The blue line illustrates the cost associated with each of the evaluated models. The arrow points at the cost of the selected model, allowing us to understand the cost relative to the theoretical range of feasible costs which are represented by the surface plot. The graph demonstrates that, despite the fact that the \ac{ml} prediction cannot perfectly anticipate all requirements, the selected cost model closely approximates the theoretical optimum.}

The traditional approach to resource acquisition involves a long term resource purchase. Therefore, $p_t = 1$ for each time step $t$. This results in a cost as proposed in equation~\eqref{eq:opt_problem} that can be calculated as follows:
\begin{equation}\label{eq:opt_problem_trad}
C_{trad} = \sum\limits_{t}c_{p_t}\cdot (1-r_t)
\end{equation}
\textcolor{corr2}{We compare this traditional approach to our adaptive resource acquisition mechanism. The comparison was done on the test set of our dataset. As shown in equation~\eqref{eq:opt_problem_trad}, traditionally we would acquire resources long term and then the cost only depends on the actual resource requirements over time, i.e. $r_t$. In contrast, our method employs the \ac{ml} models outlined in Section~\ref{sec:ml_modeling} to dynamically match the required - $r_t$ and available resources - $p_t$, achieving a cost reduction by over $80\%$ compared to the traditional approach. } 

\section{Conclusions}
In this paper we propose and study the application of ML to carry out cost optimisation based on handover predictions in 5G mobile cells, for use cases where different service providers might target different trade-offs between cost of network resource over-booking and potential lack of resources when required. 


We propose and evaluate an intelligent algorithm that predicts handover events based only on parameters measured in a commercial O-RAN system using outdoor licensed bands at 3.9 GHz, running on the OpenIreland open networking testbed infrastructure. Our results show that an \ac{lstm} model can be optimized for both recall and precision depending on the optimization objective. Specifically, in our evaluation, after setting the minimum precision to $75\%$, a recall of $88\%$ was achieved. The studied models have also been evaluated in terms of their effect on the cost of false resources acquisitions, and show a reduction in cost by more than $80\%$, when compared to a traditional long term resource purchase.

We also argue that different network stakeholders will have different objectives depending on their business models. We provide examples of conflicting optimization objectives that can be achieved on top of the same shared infrastructure by taking advantage of virtualization, slicing, functional splits and the availability of intelligent control loops proposed by the O-RAN community. 

In terms of future work on this topic, we believe that it is important to understand the applicability of federated and transfer learning techniques. Federated learning would allow us to reduce the amount of information shared through the network and to conserve privacy of information through local learning. It would also allow us to specialize the local models for the local environment covered by a small subset of \acp{bs}. Transfer learning on the other hand would allow us to speed up the training and deployment of such models in on-demand and in real time. Additionally, since the price of resources can vary over time depending on the demand, we believe that a study should be conducted in which the costs of false resource acquisitions ($c_{n_t}$ and $c_{p_t}$) are not constant either.

\section*{Acknowledgment}
This material is based upon works supported by the Science Foundation Ireland under Grants No. 17/CDA/4760, 13/RC/2077\_P2 and 18/RI/5721.

\ifCLASSOPTIONcaptionsoff
  \newpage
\fi

\bibliographystyle{IEEEtran}
\bibliography{main.bib}

\begin{thebibliography}{10}
\providecommand{\url}[1]{#1}
\csname url@samestyle\endcsname
\providecommand{\newblock}{\relax}
\providecommand{\bibinfo}[2]{#2}
\providecommand{\BIBentrySTDinterwordspacing}{\spaceskip=0pt\relax}
\providecommand{\BIBentryALTinterwordstretchfactor}{4}
\providecommand{\BIBentryALTinterwordspacing}{\spaceskip=\fontdimen2\font plus
\BIBentryALTinterwordstretchfactor\fontdimen3\font minus \fontdimen4\font\relax}
\providecommand{\BIBforeignlanguage}[2]{{%
\expandafter\ifx\csname l@#1\endcsname\relax
\typeout{** WARNING: IEEEtran.bst: No hyphenation pattern has been}%
\typeout{** loaded for the language `#1'. Using the pattern for}%
\typeout{** the default language instead.}%
\else
\language=\csname l@#1\endcsname
\fi
#2}}
\providecommand{\BIBdecl}{\relax}
\BIBdecl

\bibitem{schmidt2021flexric}
R.~Schmidt, M.~Irazabal, and N.~Nikaein, ``Flexric: an sdk for next-generation sd-rans,'' in \emph{International Conference on emerging Networking Experiments and Technologies}, 2021, pp. 411--425.

\bibitem{3gppRelease14}
\BIBentryALTinterwordspacing
3GPP, ``{Study on new Radio Access Technology: Radio Access Architecture and Interfaces},'' {3rd Generation Partnership Project (3GPP)}, Technical Specification (TS) 38.801, 2017, version 14.0.0. [Online]. Available: \url{https://portal.3gpp.org/desktopmodules/Specifications/SpecificationDetails.aspx?specificationId=3056}
\BIBentrySTDinterwordspacing

\bibitem{3gpp2019nr}
G.~T. 38.300, ``Nr; nr and ng-ran overall description; stage 2,'' 2019.

\bibitem{rodriguez2020cloud}
V.~Q. Rodriguez, F.~Guillemin, A.~Ferrieux, and L.~Thomas, ``Cloud-ran functional split for an efficient fronthaul network,'' in \emph{International Wireless Communications and Mobile Computing (IWCMC)}.\hskip 1em plus 0.5em minus 0.4em\relax IEEE, 2020, pp. 245--250.

\bibitem{duan2016performance}
J.~Duan, X.~Lagrange, and F.~Guilloud, ``Performance analysis of several functional splits in c-ran,'' in \emph{IEEE Vehicular Technology Conference}.\hskip 1em plus 0.5em minus 0.4em\relax IEEE, 2016, pp. 1--5.

\bibitem{3gpp2011evolved}
G.~T. 36.300, ``Evolved universal terrestrial radio access (e-utra) and evolved universal terrestrial radio access network (e-utran); overall description; stage 2,'' 2011.

\bibitem{polese2022understanding}
M.~Polese, L.~Bonati, S.~D'Oro, S.~Basagni, and T.~Melodia, ``Understanding o-ran: Architecture, interfaces, algorithms, security, and research challenges,'' \emph{arXiv preprint arXiv:2202.01032}, 2022.

\bibitem{oran}
\BIBentryALTinterwordspacing
O-RAN, ``{O-RAN: Towards an open and smart RAN},'' {Open Radio Access Network Alliance}, Technical Specification (TS) 38.801, 2018, online; Accessed 10-November-2022. [Online]. Available: \url{https://www.o-ran.org/resources}
\BIBentrySTDinterwordspacing

\bibitem{alliance2019ran}
O.~Alliance, ``O-ran whitepaper-building the next generation ran,'' \emph{O-RAN Alliance, Tech. Rep., Oct}, 2019.

\bibitem{alliance181017ran}
------, ``O-ran: Towards an open and smart ran.'' \emph{URL: https://www.o-ran.org/s/O-RAN-WP-FInal-181017.pdf}, 2018.

\bibitem{alliance2020ranG3}
------, ``O-ran near-real-time ran intelligent controller architecture e2 general aspects and principles 1.01 (o-ran. wg3. e2gap-v01. 01),'' \emph{Technical Specification}, 2020.

\bibitem{alliance2021ranG2}
------, ``O-ran non-rt ric: Functional architecture 1.01-march 2021 (o-ran. wg2. non-rt-ric-arch-tr-v01. 01),'' \emph{Technical Specification}, 2021.

\bibitem{alliance2019ran2}
------, ``O-ran working group 2 ai/ml workflow description and requirements,'' \emph{ORAN-WG2. AIML. v01}, vol.~1, 2019.

\bibitem{dryjanski2021toward}
M.~Dryja{\'n}ski, {\L}.~Ku{\l}acz, and A.~Kliks, ``Toward modular and flexible open ran implementations in 6g networks: Traffic steering use case and o-ran xapps,'' \emph{Sensors}, vol.~21, no.~24, p. 8173, 2021.

\bibitem{johnson2021open}
D.~Johnson, D.~Maas, and J.~Van Der~Merwe, ``Open source ran slicing on powder: A top-to-bottom o-ran use case,'' in \emph{Proceedings of the 19th Annual International Conference on Mobile Systems, Applications, and Services}, 2021, pp. 507--508.

\bibitem{bashir2019optimal}
A.~K. Bashir, R.~Arul, S.~Basheer, G.~Raja, R.~Jayaraman, and N.~M.~F. Qureshi, ``An optimal multitier resource allocation of cloud ran in 5g using machine learning,'' \emph{Transactions on emerging telecommunications technologies}, vol.~30, no.~8, p. e3627, 2019.

\bibitem{zhang2022federated}
H.~Zhang, H.~Zhou, and M.~Erol-Kantarci, ``Federated deep reinforcement learning for resource allocation in o-ran slicing,'' \emph{arXiv preprint arXiv:2208.01736}, 2022.

\bibitem{zhang2022team}
------, ``Team learning-based resource allocation for open radio access network (o-ran),'' \emph{arXiv preprint arXiv:2201.07385}, 2022.

\bibitem{9903911}
H.~Luo and H.-Y. Wei, ``Resource orchestration at the edge: Intelligent management of mmwave ran and gaming application qoe enhancement,'' \emph{IEEE Transactions on Network and Service Management}, pp. 1--1, 2022.

\bibitem{ali2021multi}
Z.~Ali, L.~Giupponi, M.~Miozzo, and P.~Dini, ``Multi-task learning for efficient management of beyond 5g radio access network architectures,'' \emph{IEEE Access}, vol.~9, pp. 158\,892--158\,907, 2021.

\bibitem{dzaferagic2022Globe}
M.~Dzaferagic, J.~Ayala-Romero, and M.~Ruffini, ``{ML Approach for Power Consumption Prediction in Virtualized Base Stations},'' \emph{IEEE Globecom}, 2022.

\bibitem{bhattarai2016overview}
S.~Bhattarai, J.-M.~J. Park, B.~Gao, K.~Bian, and W.~Lehr, ``An overview of dynamic spectrum sharing: Ongoing initiatives, challenges, and a roadmap for future research,'' \emph{IEEE Transactions on Cognitive Communications and Networking}, vol.~2, no.~2, pp. 110--128, 2016.

\bibitem{ansari2020spectrum}
R.~I. Ansari, N.~Ashraf, S.~A. Hassan, G.~Deepak, H.~Pervaiz, and C.~Politis, ``Spectrum on demand: a competitive open market model for spectrum sharing for uav-assisted communications,'' \emph{IEEE Network}, vol.~34, no.~6, pp. 318--324, 2020.

\bibitem{roy2004exploiting}
A.~Roy, S.~K. Das, and A.~Misra, ``Exploiting information theory for adaptive mobility and resource management in future cellular networks,'' \emph{IEEE Wireless Communications}, vol.~11, no.~4, pp. 59--65, 2004.

\bibitem{demissie2013exploring}
M.~G. Demissie, G.~H. de~Almeida~Correia, and C.~Bento, ``Exploring cellular network handover information for urban mobility analysis,'' \emph{Journal of Transport Geography}, vol.~31, pp. 164--170, 2013.

\bibitem{yan2021mobility}
M.~Yan, S.~Li, C.~A. Chan, Y.~Shen, and Y.~Yu, ``Mobility prediction using a weighted markov model based on mobile user classification,'' \emph{Sensors}, vol.~21, no.~5, p. 1740, 2021.

\bibitem{9171421}
M.~Dzaferagic, N.~Marchetti, and I.~Macaluso, ``Minimizing the signaling overhead and latency based on users’ mobility patterns,'' \emph{IEEE Systems Journal}, vol.~15, no.~1, pp. 77--84, 2021.

\bibitem{sun2020efficient}
Y.~Sun, W.~Jiang, G.~Feng, P.~V. Klaine, L.~Zhang, M.~A. Imran, and Y.-C. Liang, ``Efficient handover mechanism for radio access network slicing by exploiting distributed learning,'' \emph{IEEE Transactions on Network and Service Management}, vol.~17, no.~4, pp. 2620--2633, 2020.

\bibitem{kaur2022efficient}
G.~Kaur, R.~K. Goyal, and R.~Mehta, ``An efficient handover mechanism for 5g networks using hybridization of lstm and svm,'' \emph{Multimedia Tools and Applications}, vol.~81, no.~26, pp. 37\,057--37\,085, 2022.

\bibitem{cai2021mobility}
Y.~Cai, Y.~Chen, M.~Ding, P.~Cheng, and J.~Li, ``Mobility prediction-based wireless edge caching using deep reinforcement learning,'' in \emph{2021 IEEE/CIC International Conference on Communications in China (ICCC)}.\hskip 1em plus 0.5em minus 0.4em\relax IEEE, 2021, pp. 1036--1041.

\bibitem{AKYILDIZ201617}
``5g roadmap: 10 key enabling technologies,'' \emph{Computer Networks}, vol. 106, pp. 17--48, 2016.

\bibitem{farooq2019impact}
M.~Farooq and V.~Raju, ``Impact of over-the-top (ott) services on the telecom companies in the era of transformative marketing,'' \emph{Global Journal of Flexible Systems Management}, vol.~20, no.~2, pp. 177--188, 2019.

\bibitem{e2sm-kpm}
{O-RAN Alliance WG3}, ``{O-RAN Near-Real-time RAN Intelligent Controller E2 Service Model (E2SM) KPM 2.0},'' Tech. Rep., 2021.

\bibitem{3gpp_38_913_v_14_2_r_14}
{3GPP}, ``{TR 38.913 Version 14.2.0 Release 14},'' Tech. Rep., 2017.

\end{thebibliography}

\end{document}